\documentclass{iopjournal}

\usepackage{graphicx}%
\usepackage{multirow}%
\usepackage{amsmath,amssymb,amsfonts}%
\usepackage{amsthm}%
\usepackage{mathrsfs}%
\usepackage[title]{appendix}%
\usepackage{xcolor}%
\usepackage{textcomp}%
\usepackage{manyfoot}%
\usepackage{booktabs}%
\usepackage{orcidlink}
\usepackage{comment}
\usepackage{algorithm}%
\usepackage{algorithmicx}%
\usepackage{algpseudocode}%
\usepackage{listings}%
\usepackage{makecell}%
\usepackage{ragged2e}
\usepackage{cite}

\begin{document}

\articletype{Paper} 

\title{Phenomenological Noise Models and Optimal Thresholds of the 3D Toric Code}

\author{
Ji-Ze Xu$^{1,2}$\orcid{0009-0009-4831-0933}, 
Yin Zhong$^{3,4}$\orcid{0000-0003-3748-9973}, 
Miguel A. Martin-Delgado$^{5,6}$\orcid{0000-0003-2746-5062}
Hao Song$^{7,*}$\orcid{0000-0001-5513-8857}
and Ke Liu$^{1,8,\dagger}$\orcid{0000-0002-5768-1631}
}

\affil{$^1$Hefei National Research Center for Physical Sciences at the Microscale and School of Physical Sciences, University of Science and Technology of China, Hefei, 230026, China}

\affil{$^2$Hefei National Laboratory, University of Science and Technology of China, Hefei, 230088, China}

\affil{$^3$Key Laboratory of Quantum Theory and Applications of MoE $\&$ School of Physical Science and Technology, Lanzhou University, Lanzhou, 730000, China}

\affil{$^4$Lanzhou Center for Theoretical Physics $\&$  Key Laboratory of Theoretical Physics of Gansu Province, Lanzhou University, Lanzhou, 730000, China}

\affil{$^5$Departamento de F\'isica Te\'orica, Universidad Complutense, Madrid, 28040, Spain}

\affil{$^6$CCS-Center for Computational Simulation, Universidad Polit\'ecnica de Madrid, Madrid, 28660, Spain}

\affil{$^7$ Institute of Theoretical Physics, Chinese Academy of Sciences, Beijing, 100190, China}

\affil{$^8$Shanghai Research Center for Quantum Science and CAS Center for Excellence in Quantum Information and Quantum Physics, University of Science and Technology of China, Shanghai, 201315, China}

\affil{$^{*,\dagger}$Author to whom any correspondence should be addressed.}

\email{songhao@itp.ac.cn, ke.liu@ustc.edu.cn}


\justifying

\begin{abstract}
\justifying
Three-dimensional (3D) topological codes offer the advantage of supporting fault-tolerant implementations of non-Clifford gates, yet their performance against realistic noise remains largely unexplored. 
In this work, we focus on the paradigmatic 3D toric code and investigate its fault-tolerance thresholds in the presence of both Pauli and measurement errors. 
Two randomly coupled lattice gauge models that describe the code's correctability are derived, including a random 2-form $\mathbb{Z}_2$ gauge theory. 
By exploiting a generalized duality technique, we show that the 3D toric code exhibits optimal thresholds of $p^{X,M}_{th} \approx 11\%$ and $p^{Z,M}_{th} \approx 2\%$ against bit-flip and phase-flip errors, respectively. 
These threshold values show modest reductions compared to the case of perfect measurements, establishing the robustness of the 3D toric code against measurement errors. 
Our results constitute a substantial advance towards assessing the practical performance of 3D topological codes. 
This contribution is timely and in high demand, as rapid hardware advancements are bringing complex codes into experimental reach. 
Moreover, our work highlights the interdisciplinary nature of fault-tolerant quantum computation and holds significant interest for quantum information science, high-energy physics, and condensed matter physics.
\end{abstract}

\section{Introduction}\label{sec:intro}

The realization of scalable quantum computation hinges on quantum error correction (QEC) to protect quantum information from noise\cite{Dennis02, Fowler12, Shor95, Steane96, Kitaev03}.
For years, the primary candidates for implementing QEC have been 2D topological codes, such as the 2D toric code~\cite{Kitaev03,Bravyi98}, color code~\cite{Bombin06}, and their variants. 
These codes are favored for their local stabilizer structures and high fault-tolerance thresholds, and have been successfully demonstrated on various quantum computing platforms~\cite{Satzinger21,Krinner22,USTC22,USTC23,Google24a,Google25_color,Quantinuum24,QuEra24,QuEra25,Monz21,Blatt14}.
However, the utility of these 2D codes is fundamentally constrained by the Bravyi-Koenig theorem~\cite{Bravyi13}, which forbids the fault-tolerant implementation of non-Clifford gates.
Consequently, to achieve universal computation, one has to resort to protocols like magic state distillation (MSD)~\cite{Bravyi05}, which are notoriously resource-intensive.
A similar situation is encountered by prominent quantum low-density parity-check (qLDPC) codes, including bivariate-bicycle codes~\cite{Bravyi24} and hypergraph product codes~\cite{Tillich14}. 
Despite these codes offering a high encoding rate for efficient quantum memory, their hardware relevant realizations also rely on MSD to perform non-Clifford gates~\cite{Xu25,Yoder25}.
In stark contrast, 3D topological codes can directly implement non-Clifford gates in a fault-tolerant manner, such as the logical T gate in the 3D color code~\cite{Bombin07} and the logical CCZ gate in the 3D toric code~\cite{Vasmer19,Barkeshli24}.
They thus offer a compelling pathway toward scalable and universal fault-tolerant gate sets, and are becoming increasingly viable given recent rapid progress in platforms like reconfigurable atom arrays and highly-connected trapped-ion systems.

Nevertheless, despite their computational advantages, a comprehensive understanding of the performance of 3D topological codes remains elusive. 
This is largely due to two related challenges: the lack of efficient decoders and an incomplete knowledge of their fault-tolerant thresholds. 
Decoding 3D codes is notoriously difficult due to the complexity of their error and syndrome spaces and the hypergraph nature of the underlying problem.
Existing decoders are typically either limited to small code sizes, such as BP-OSD based decoders~\cite{Panteleev21}, or can only decode parts of errors that are compatible with single-shot or matchable algorithms~\cite{Quintavalle21,Brown16}. 
Thus, they cannot yet reflect the true performance of a 3D code.

A powerful alternative for determining a code's potential is the statistical-mechanical mapping method introduced in the seminal paper Ref.~\cite{Dennis02}, which recasts the threshold problem as a phase transition in a random spin model.
This approach is decoder-independent and can compute the optimal threshold of a code. 
However, although it has been widely used to analyze 2D codes~\cite{Dennis02, Bombin12,Chubb21,Fan24,fx56-8nvy,gskb-t5ql}, its applications to 3D codes have mostly been restricted to the simplest error models that assume perfect syndrome measurements~\cite{Wang03,Kubica18,Song22,Canossa26}.
In any realistic device, measurement errors are inevitable and are known to degrade thresholds.
The inclusion of these errors not only adds an effective time dimension but also fundamentally alters the nature of the underlying interactions, dramatically complicating both the derivation of the statistical-mechanical model and the analysis of its phase transition.
The resulting complexity has thus hindered systematic assessment of the practical performance of 3D codes, either as the primary code in a QEC scheme or as fault-tolerance gadgets in a hybrid architecture~\cite{Beverland21,Wang24,Xu25b}.

In this work, we focus on the 3D toric code, a paradigmatic 3D topological code, and investigate the optimal phenomenological thresholds.
We derive the effective random spin models that describe the code's correctability subject to both stochastic Pauli and measurement errors.
These models include the well-known 4D random Ising gauge theory~\cite{Nishimori07} and a novel 4D random 2-form gauge model defined by face variables and cube interactions. 
We analyze their gauge-invariant order parameters and confinement-deconfinement phase transitions, in particular on the Wilson sheet operator and the spontaneous breaking of 2-form symmetry in the latter model. 
The thresholds are computed analytically by employing duality techniques that lead to a generalized entropy relation incorporating quenched disorders. 
This allows us to bypass direct numerical simulations of these 4D random models, which are notoriously resource-intensive.

We find that the 3D toric code, when subjected to measurement errors, exhibits a bit-flip error threshold of $p^{X,M}_{th} \approx 0.11$~\cite{Nishimori07}  and a phase-flip error threshold $p^{Z,M}_{th} \approx 0.02$, reduced from their respective perfect measurement thresholds $p^X_{th} \approx 0.23$ and $p^Z_{th} \approx 0.11$. 
The code's overall optimal phenomenological threshold is set by the lower value, hence $p^{Z,M}_{th} \approx 0.02$.
Notably, this reduction is modest compared to the 2D toric code, whose threshold drops more sharply from $0.11$ to $0.033$ due to measurement errors~\cite{Wang03}.
Furthermore, the optimal bit-flip threshold $p^{X,M}_{th} \approx 0.11$ is considerably higher than the value $0.071$ found by a neural network decoder~\cite{Breuckmann18} and the value $0.029$ found by single-shot decoders~\cite{Quintavalle21, note_threshold}, while a recently developed overlapping window decoder has found a threshold of $0.0965$~\cite{Hillmann25} close to the optimal value.
For the phase-flip error threshold, $p^{Z,M}_{th} \approx 0.02$, the only comparable result in the literature is $0.0126$ computed recently by a matching decoder~\cite{Berent24}. 
This indicates a significant opportunity to improve 3D code decoding algorithms.

Our work establishes the robustness of the 3D toric code against faulty measurements, thus forging a critical link between theoretical analysis and the realistic conditions in quantum devices.
It also presents the first calculation of the optimal error thresholds for 3D topological codes in the presence of measurement errors, a substantial advance over previous studies that assumed perfect measurements.
Given the preeminent representative role of the toric code, our results constitute a reference point for evaluating the practical performance of 3D topological codes.

Moreover, our derivations of the 2-form gauge theory and the utilization of duality techniques deepen the connection among quantum information science, high-energy physics, and statistical physics. 
Going beyond the well-known link between the 2D toric code and the standard Ising lattice gauge theory, our findings suggest that more complex topological codes offer a rich landscape for constructing novel gauge theories.
This can further stimulate studies into the associated topological states of matter and their phase transitions from condensed matter and statistical-mechanical perspectives. 
In turn, new physics and techniques developed within these disciplines can serve to elucidate the mechanisms and provide powerful tools for analyzing intricate quantum error-correcting codes.

\section{Preliminaries}

We begin by defining the 3D toric code and the basics of quantum error correction.

\subsection{The 3D toric code}

\begin{figure}[tb!]
	\centering
	\includegraphics[width=0.7\columnwidth]{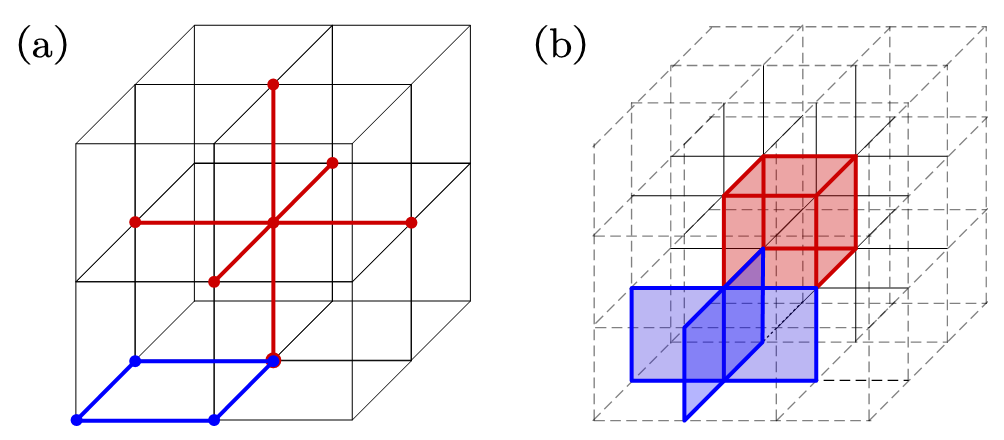}
	\caption{\textbf{Stabilizers of the 3D toric code.} 
(a) Each link of the lattice hosts one physical qubit. 
The vertex operator $A_s$ (red) acts on the six qubits located on the links incident to vertex $s$. 
The plaquette operator $B_p$ (blue) acts on the four qubits surrounding plaquette $p$. (b) The same operators viewed on the dual lattice.}
	\label{stabilizers}
\end{figure}

The 3D toric code is defined on an $L \times L \times L$ cubic lattice with periodic boundary conditions, with a physical qubit residing on each link. It is constructed from two types of commuting stabilizers: vertex operators ($A_s$) and plaquette operators ($B_p$):
\begin{equation}
	A_s=\prod _{\ell \in \text{star}(s)}X_{\ell},\qquad B_p=\prod _{\ell\in \partial p}Z_{\ell},
\end{equation}
where $A_s$ is the product of Pauli-$X$ operators on the star of links incident on a vertex $s$, and $B_p$ is the product of Pauli-$Z$ operators on the links forming the boundary of a plaquette $p$, as illustrated in Fig.~\ref{stabilizers}(a). Sometimes, it is convenient to utilize the dual lattice, where qubits are labeled by dual plaquettes (Fig.~\ref{stabilizers}(b)).

The degenerate ground states of the Hamiltonian $H=-\sum_{s}A_s-\sum_{p}B_p$ define the code space: the subspace where all stabilizers have an eigenvalue of $+1$. Errors excite the system out of this subspace, creating quasiparticle excitations where stabilizers are violated (yield an eigenvalue of $-1$). Violations of $A_s$ and $B_p$ stabilizers correspond to point-like ``electric charges'' ($e$) at vertices and loop-like ``magnetic fluxes'' ($m$) on plaquettes, respectively. For convenience, magnetic fluxes are often analyzed on the dual lattice, where they appear as excitations on dual links.

\begin{figure}[tb!]
	\centering
	\includegraphics[scale=0.7]{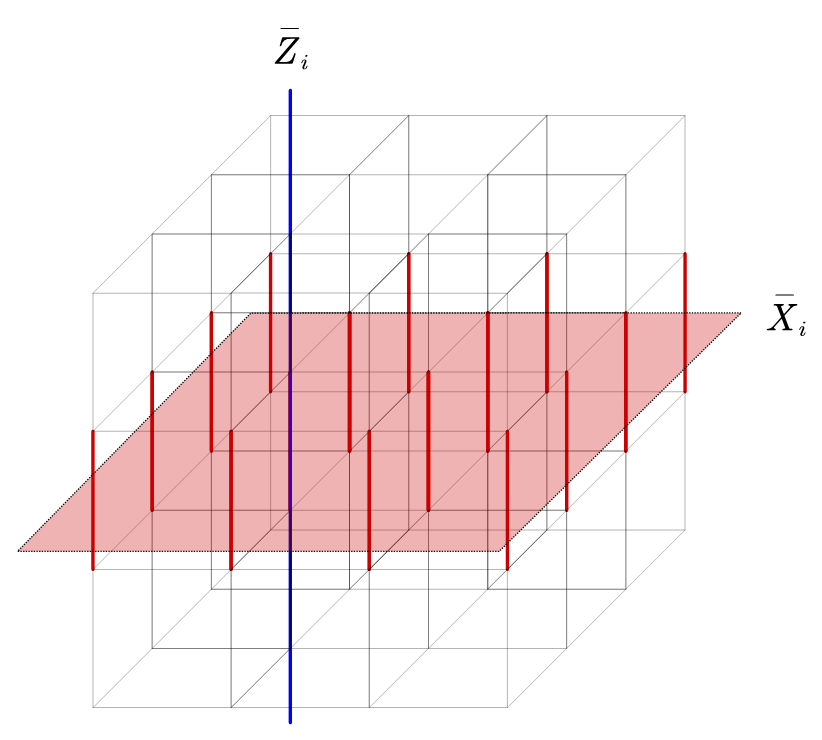}
	\caption{\textbf{Logical operators of the 3D toric code.} 
The logical operator $\bar{Z}_i$ (blue) is a product of Pauli-$Z$ operators along a noncontractible string that winds around the lattice. 
The logical operator $\bar{X}_i$ (red) is a product of Pauli-$X$ operators acting on the links intersected by a noncontractible membrane on the dual lattice. 
These two logical operators anticommute because the string and membrane intersect at a single qubit.}
	\label{logical_operators}
\end{figure}

The code encodes $k=3$ logical qubits. The logical Pauli operators, $\bar{Z}_i$ and $\bar{X}_i$ for $i \in \{1, 2, 3\}$, have a clear geometric interpretation as non-trivial operators that commute with all stabilizers:
\begin{equation}
	\bar{Z}_i=\prod _{\ell\in \mathcal{L}_i}Z_{\ell}, \qquad \bar{X}_i=\prod _{p^*\in \mathcal{P}_i^*}X_{p^*}.
\end{equation}
Here, $\mathcal{L}_i$ is a non-trivial closed string of links wrapping a handle of the torus, while $\mathcal{P}_i^*$ is a non-trivial closed membrane of plaquettes on the dual lattice, wrapping a dual handle (Fig.~\ref{logical_operators}). These operators satisfy the Pauli algebra for three independent qubits.

\subsection{Error correction}

The 3D toric code is a Calderbank-Shor-Steane (CSS) code, thus its $X$ and $Z$ errors can be analyzed separately.
Consider random bit-flip ($X$) and phase-flip ($Z$) errors that occur independently on each qubit with probability $p$. Errors are detected by measuring the stabilizers, and the resulting set of violated stabilizers is the syndrome, which pinpoints the locations of quasiparticle excitations.

\begin{figure}[tb!]
	\centering
	\includegraphics[width=0.8\linewidth]{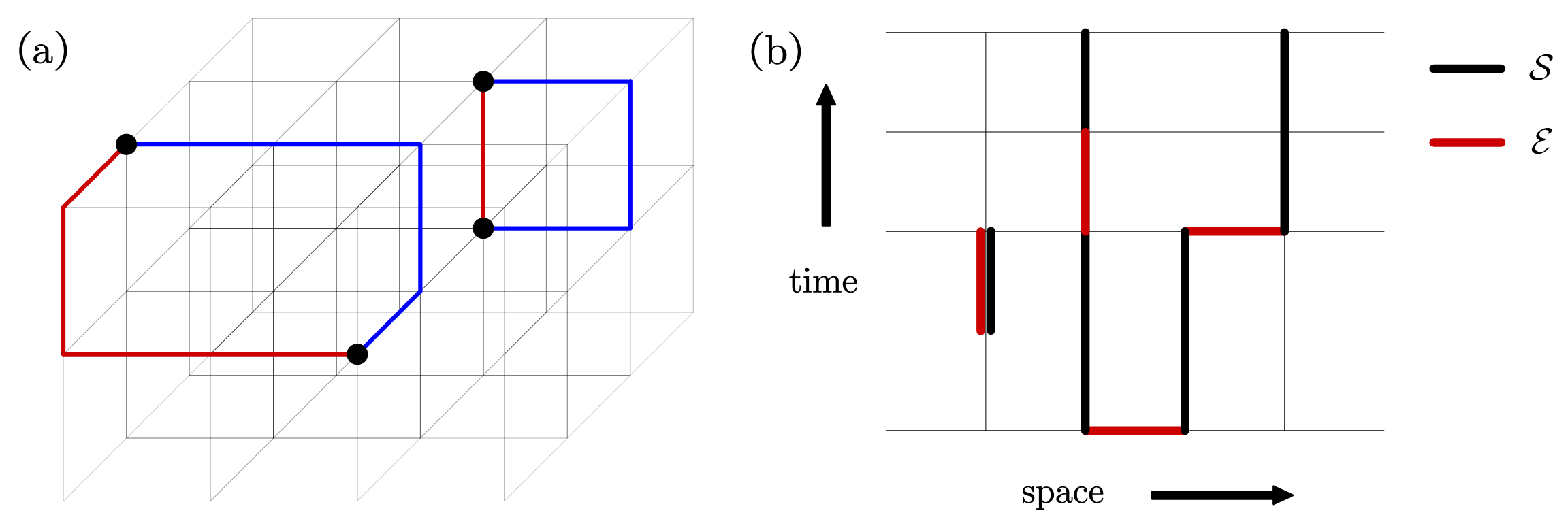}
	\caption{\textbf{Geometric picture of errors and syndromes.} 
(a) A spatial lattice showing a syndrome configuration (black dots) and two possible $Z$-error chains, $\mathcal{E}$ (red links) and $\mathcal{E}'$ (blue links), that both lead to this syndrome. 
(b) A spacetime lattice showing a syndrome configuration $\mathcal{S}$ (black temporal links) obtained from faulty measurements of the $A$-type stabilizers, together with a compatible error chain $\mathcal{E}$ that includes both $Z$ errors (red spatial links) and measurement faults (red temporal links).
    }
	\label{fig_error_syndrome}
\end{figure}

To formalize the relationship between an error and its syndrome, we employ the language of homology. An error is represented as a chain—a collection of links for $Z$ errors or dual plaquettes for $X$ errors—and the syndrome is precisely the boundary of this error chain. This relationship is captured by a chain complex. For phase-flip errors, the complex is:
\begin{equation}
	\mathcal{G}_z \xrightarrow{\partial} \mathcal{E} \xrightarrow{\partial} \mathcal{G}_x,
\end{equation}
where $\mathcal{E}$ is the 1-chain of link errors and its boundary $\partial\mathcal{E}$ is the 0-chain of vertex syndromes. For bit-flip errors, it is convenient to use the dual lattice (Fig.~\ref{stabilizers}(b)), and the corresponding dual chain complex is:
\begin{equation}
	\mathcal{G}_z^* \xleftarrow{\partial^*} \mathcal{E}^* \xleftarrow{\partial^*} \mathcal{G}_x^*.
\end{equation} 
Two errors, $\mathcal{E}$ and $\mathcal{E}'$, are in the same class if their sum is a homologically trivial cycle (i.e., a boundary) as in Fig.~\ref{fig_error_syndrome}(a), which corresponds to a product of stabilizers. Different homology classes are separated by homologically non-trivial cycles, which are the logical operators $\mathcal{L}$. Therefore, any error $\mathcal{E}'$ in a different class can be written as $\mathcal{E}' = \mathcal{E} + \mathcal{L}$.
The decoder's task is to infer the most likely error equivalence class given the measured syndrome. 

This framework robustly extends to faulty syndrome measurements by analyzing the process on a 4D spacetime lattice, where time represents repeated measurement cycles. As illustrated in Fig.~\ref{fig_error_syndrome}(b), the total error chain $\mathcal{E}$ (including qubit and measurement errors) and the syndrome history $\mathcal{S}$ are now chains in this 4D space. 
To correct the encoded quantum information, one must be able to infer the correct homology class of the total error $\mathcal{E}$ from the syndrome history $\mathcal{S}$. 
For $p < p_{th}$, the probability of this inference failing vanishes in the thermodynamic limit, guaranteeing fault-tolerant correction.

\section{Methods}\label{sec:method}

Here we outline the methods employed in this work, while detailed derivations are provided in Supplementary material~\cite{SM}. 

\subsection{Homological formalism for error correction}

We consider both random single-qubit Pauli errors and measurement errors.
A single qubit has three noise channels, corresponding to $X$, $Y$, and $Z$ errors.
Only two of them are independent, as a $Y$ error can be decomposed into $\mathrm{i}XZ$.
In addition, since the 3D toric code is a CSS code, we can analyze the effects of the $X$ and $Z$ errors separately.

First consider the case without measurement errors. An arbitrary $X$ or $Z$ error channel can be expressed as
\begin{equation}
	E _Z=\prod _{\ell\in\mathcal{E} _Z}Z_{\ell},\qquad E _X=\prod _{\ell\in\mathcal{E} _X}X_{\ell},
\end{equation}
where $\mathcal{E} _Z$ ($\mathcal{E} _X$) is the set of all links where the qubits experience $Z$ ($X$) errors. 
The destruction of the encoded state is described by
\begin{equation}
	\begin{aligned}
		\rho &\to\rho^\prime_Z=E _Z\rho E _Z^ \dagger \\
		\rho &\to\rho^\prime_X=E _X\rho E _X^ \dagger,
	\end{aligned}
\end{equation}
where $\rho$ and $\rho^\prime$ denote the density matrix before and after the error, respectively.
The goal of error correction is to find a recovery operator $R$ for the corrupted state $\rho^\prime$ such that $R\rho^\prime R ^ \dagger=\rho $, which must satisfy $ER = g$, where $g$ is a stabilizer.
In other words, the combination of $R$ and $E$ shall form trivial loops.
For simplicity, we have omitted the $Z$ or $X$ subscript provided there is no confusion.

The above picture can be extended to the situation with measurement errors, by introducing an extra temporal dimension due to repeated syndrome measurements, as seen in Fig.~\ref{fig_error_syndrome}.
Error chains and recovery operators are then defined on a 4D spacetime lattice, where timelike links represent measurement errors, and spacelike links represent qubit errors.
In the homological formalism, the errors, measurement syndrome, and recovery are represented as $k$-chains in a chain complex. 
Accordingly, successful error correction can be expressed as the homological condition 
\begin{equation}
	\mathcal{E} +\mathcal{S} +\mathcal{R} =\mathcal{C},
\end{equation}
where $\mathcal{E}$ denotes the total error chain, including both qubit and measurement errors, $\mathcal{S}$ is the measurement syndrome chain, $\mathcal{R}$ is the applied recovery chain, and $\mathcal{C}$ is any homologically trivial cycle.

The error chain $\mathcal{E}$ that results in the same syndrome chain $\mathcal{S}$ is not unique, and all such error chains together form a homology class $\bar{\mathcal{E}} = \mathcal{E} + \{\mathcal{C}\}$. In practice, the decoder's goal is to derive the equivalence class $\bar{\mathcal{E} }$ from the measurement syndrome $\mathcal{S}$, so that errors are corrected by the recovery chain $\mathcal{R}\in\bar{\mathcal{E} } + \mathcal{S}$. In contrast, if decoding yields an incorrect equivalence class $\bar{\mathcal{E}}_\mathcal{L} = \bar{\mathcal{E} } + \mathcal{L}$, where $\mathcal{L}$ is a nontrivial cycle carrying a logical operation, the recovery chain $\mathcal{R}\in\bar{\mathcal{E}}_\mathcal{L} + \mathcal{S}$ results in the failure of error correction. For topological codes, there exists an error threshold $p_{th}$ below which the errors are always correctable in the large size limit. Formally, this means that the probability of misidentifying the error class vanishes, i.e.,
\begin{equation}
	\lim_{L\to \infty}\frac{{\rm pr}(\bar{\mathcal{E} }_\mathcal{L} |\mathcal{S} )}{{\rm pr}(\bar{\mathcal{E} }|\mathcal{S} )}\to 0 ,\label{th_condition}
\end{equation}
where ${\rm pr}(\bar{\mathcal{E} }|\mathcal{S} )$ is the conditional probability that the decoding result is $\bar{\mathcal{E} }$ when the syndrome is $\mathcal{S}$, and $L$ is the linear size of the lattice.

\begin{figure}[ht!]
	\centering
	\includegraphics[width=0.5\columnwidth]{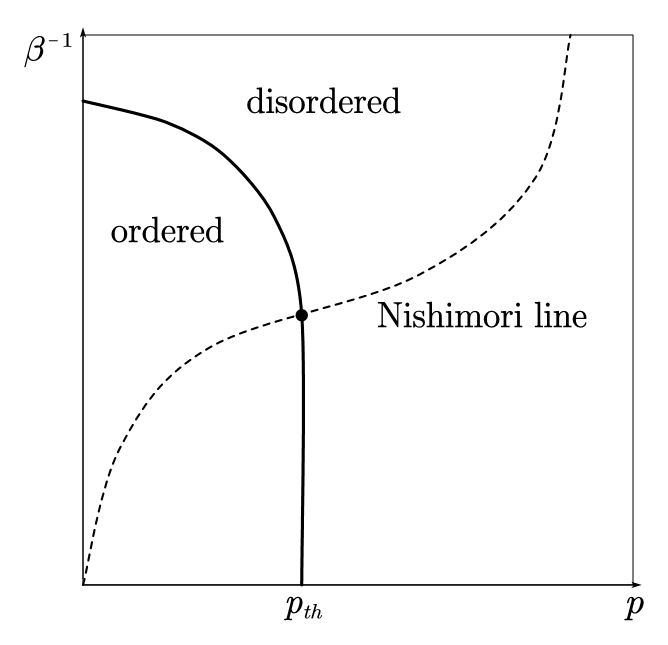}
	\caption{\textbf{Schematic phase diagram and the Nishimori line for error correction.} The correctability of an error-correcting code is mapped to the phase diagram of a statistical-mechanical model. The horizontal axis is the physical error rate $p$, and the vertical axis is inverse temperature. The solid line separates the ordered (correctable) phase from disordered (uncorrectable) phases. The optimal error-correction threshold, $p_{th}$, is determined by the intersection of the Nishimori line (dotted curve) and the phase boundary.}
	\label{phase_diagram}
\end{figure}

\subsection{Statistical-mechanical mapping}\label{subsec:method2}

To determine the error-correction threshold, we construct a mapping from the topological code to statistical-mechanical models with quenched disorder. 
The establishment of the mapping is based on Eq.~\eqref{th_condition}, where the conditional probability ${\rm pr}(\bar{\mathcal{E} }|\mathcal{S} )$ can be expressed as~\cite{Dennis02}
\begin{equation}\label{eq:pr}
	\begin{aligned}
		{\rm pr}(\bar{\mathcal{E} }|\mathcal{S})=&\left[\prod_{\ell}\sqrt{p(1-p)}\right]\cdot\sum_{\{u_{\ell}\}}\prod_{\ell}\exp\big(\beta J\eta_{\ell}u_{\ell}\big) \\
		\propto&\sum_{\{u_{\ell}\}}\exp(-\beta H),
	\end{aligned}
\end{equation}
with $H=-J\sum_{\ell}\eta_{\ell}u_{\ell}(\sigma_i)$ being the formal Hamiltonian of Ising spins $\sigma_i \pm 1$.

The error rate $p$ is related to the temperature through the Nishimori relation $e^{-2\beta J} = p/(1-p)$.
The quenched disorder $\{\eta_{\ell}\}$ represents the error configuration with the probability distribution
\begin{equation}
	P(\eta_{\ell})=p\delta (\eta_{\ell}+1)+(1-p)\delta (\eta_{\ell}-1),
\end{equation}
while the spin interaction $u_{\ell}(\sigma_i)$ is determined by the specific topological code and error model.

Therefore, Eq.~\eqref{th_condition} can be written as
\begin{equation}
	\lim_{L\to \infty}\frac{{\rm pr}(\bar{\mathcal{E} }_\mathcal{L} |\mathcal{S} )}{{\rm pr}(\bar{\mathcal{E} }|\mathcal{S} )}=\frac{Z_L[J,\eta]}{Z[J,\eta]}=\exp[-\beta \delta F(J,\eta)]\to0,
\end{equation}
where $Z[J,\eta]$ is the partition function, $\delta F(J,\eta)=F_L(J,\eta)-F(J,\eta)$ is the free energy difference between the homology classes. To satisfy the above relation, the free energy difference $\delta F(J,\eta)$ diverges when $p < p_{th}$ and converges when $p > p_{th}$, corresponding to correctable and uncorrectable errors, respectively. In the limit of $L\to \infty$, the length of the nontrivial cycle $\mathcal{L}$ tends to infinity, leading to the divergence of the free energy $F_L(J,\eta)$, which in turn causes $F(J,\eta)$ to converge for $p<p_{th}$ and diverge for $p>p_{th}$.
Consequently, the error correction threshold corresponds to the critical point of the effective spin model along the Nishimori line, as shown in Fig.~\ref{phase_diagram}.

For simplicity, we have assumed qubits and measurements have the same error rate.
Thus, the phase transition along the Nishimori line can be controlled by a single parameter $p$, as $e^{-2\beta J} = p/(1-p)$.
Nonetheless, we can also introduce an independent measurement error rate, $q$, and define another coupling strength $K$, satisfying $e^{-2\beta K} = q/(1-q)$ along the Nishimori line. 
The statistical-mechanical approach also applies to general cases $p\neq q$, while the phase transition would be described by two parameters due to anisotropic coupling $J\neq K$. 
While the error threshold generally depends on the specific combination of $(p,q)$, the symmetric case $p=q$ is a representative and commonly adopted in the literature.

\subsection{Duality method and threshold determination}

The standard way to determine optimal thresholds is through numerical simulations.
It would be extremely resource intensive for the current problem due to quenched disorder and high dimensionality.
Nevertheless, this computational cost can be significantly reduced or even bypassed using duality techniques. 

To illustrate the method, we first recall the standard Kramers-Wannier duality, which applies to the disorder-free limit ($p = 0$) of our spin models. 
This duality refers to the fact that, for a classical spin model $H = - J\sum_{\ell} \prod_i \sigma_i$, there exists a dual model $\widetilde{H} = - \tilde{J}\sum_{\ell} \prod_{\tilde{i}} \tilde{\sigma}_{\tilde{i}}$ that satisfies~\cite{Wegner71} 
\begin{equation}
	\sinh(2\beta J)\sinh(2\tilde{\beta }\tilde{J})=1.
\end{equation}
This duality relation also applies to the critical points $\beta_c$ and $\tilde{\beta}_c$ of the two models.

In the presence of disorders ($p\neq 0$), the exact duality no longer holds. However, a generalized duality relation can be derived for the configuration-averaged free energy specifically along the Nishimori line. To calculate the quenched free energy, we employ the replica method, which involves computing the average of the $n$-th power of the partition function, $[Z^n]$, and then taking the analytic continuation to the limit $n \to 0$. Applying the replica trick to the partition functions of a model and its dual, we find that while the exact duality is broken, the replicated partition functions $[Z^n]$ and $[\widetilde{Z}^n]$ remain connected.

By comparing the functional forms of the replicated partition functions in the limit $n \to 0$, one can derive the generalized duality relation~\cite{Nishimori07, Song22,SM},  
\begin{equation}
	\mathcal{H}(p_{c}) + \mathcal{H}(\tilde{p}_{c}) \approx 1, \label{eq:duality}
\end{equation}
Here, $\mathcal{H}(p) = -p\log_2(p) - (1-p)\log_2(1-p)$ is the Shannon entropy, while $p_{c}$ and $\tilde{p}_{c}$ denote the critical points of the two dual spin models along the Nishimori line.

This generalized duality relation has been extensively validated. Comparisons with existing large-scale Monte Carlo simulations show that the Shannon entropies of dual pairs consistently satisfy $\mathcal{H}(p_c)+\mathcal{H}(\tilde{p}_c)\approx 1.00(2)$, with only minor deviations from the ideal duality value of 1. Examples include various random spin models~\cite{Nishimori07}, topological codes~\cite{Wang03,Bombin12,Kubica18}, and fracton codes~\cite{Song22,Canossa26}.
In present work, the generalized duality relation will allow us to determine the previously unknown phenomenological threshold for the 3D toric code via a more tractable dual description.


\section{Effective random spin models}
We now give the explicit spin Hamiltonians modeling the correctability of the 3D tori code for both perfect and imperfect measurements.
Details of the derivation are provided in SM~\cite{SM}.

The probability of an error configuration  $\mathcal{E}$ can be expressed as a Boltzmann weight for a classical spin model with quenched disorder. The probability of an error chain $\mathcal{E}' = \mathcal{E} + \mathcal{C}$ (where $\mathcal{C}$ is a homologically trivial cycle) is proportional to $\exp(-\beta H)$, with the Hamiltonian
\begin{equation}
	H = -J \sum_{\ell} \eta_{\ell} u_{\ell}.
\end{equation}
Here, $\{\eta_{\ell}\}$ represents the quenched disorder from the physical error chain $\mathcal{E}$, and $\{u_{\ell}\}$ represents the dynamic spin variables corresponding to the cycle $\mathcal{C}$. This mapping is exact when the physical error rate $p$ is related to the model's parameters by the Nishimori line: $e^{-2\beta J} = p/(1-p)$.

The probability of a given error homology class is proportional to the partition function $Z[J, \eta]$ of this model. Successful error correction depends on the ratio of partition functions for different homology classes, which is governed by the free energy difference between them. This leads to a direct correspondence where the ordered phase of the model is the correctable regime, as the free energy cost of non-trivial cycles diverges, suppressing logical errors. Conversely, the disordered phase is the uncorrectable regime, where the free energy of all large cycles diverges, making different error classes indistinguishable. The error threshold $p_c$ is the critical point where the Nishimori line intersects the phase boundary of the averaged random statistical model, as shown in Fig.~\ref{phase_diagram}.

\textbf{Perfect Measurements (3D Models).}
When syndrome measurements are perfect, the error correction problem is confined to the 3D spatial lattice.

For bit-flip ($X$) errors, the error chains are 2-chains on the 3D dual lattice. The constraint that the dynamic variables $\{u_{p^*}\}$ form a 2-cycle, $\prod_{\ell^* \in \partial p^*} u_{p^*} = 1$, is solved by defining spins $\{\sigma_s\}$ on the vertices of the original lattice, such that $u_{\ell} = \sigma_i \sigma_j$ for a link $\ell = \langle ij \rangle$. This maps the problem to the 3D random-bond Ising model (RBIM)~\cite{Wang03}, with the Hamiltonian
\begin{equation}
	H^X = -J \sum_{\langle ij \rangle} \eta_{ij} \sigma_i \sigma_j,
\end{equation}
where the random couplings $\eta_{ij}$ reflect the error configuration. The ordered (ferromagnetic) phase of the RBIM is characterized by a non-zero average magnetization, $m = \frac{1}{N} \sum_i [\langle \sigma_i \rangle]$. In this phase, the formation of large ``domain walls'' (thermal fluctuations) is suppressed, which is dual to the suppression of logical errors. The disordered (paramagnetic) phase, where $m=0$, corresponds to the uncorrectable regime.

For phase-flip ($Z$) errors, the error chains are 1-chains on the original lattice. The 1-cycle constraint, $\prod_{\ell \in \partial s} u_{\ell} = 1$, is solved by defining spins $\{\sigma_{\ell^*}\}$ on the links of the dual lattice, such that $u_{p^*} = \prod_{\ell^* \in \partial p^*} \sigma_{\ell^*}$. This leads to the 3D random-plaquette $\mathbb{Z}_2$ gauge model (RPGM), with the Hamiltonian
\begin{equation}
	H^Z = -J \sum_{p^*} \eta_{p^*} \prod_{\ell^* \in \partial p^*} \sigma_{\ell^*}.
\end{equation}
The phase transition in the RPGM is a confinement-deconfinement transition, characterized by the Wilson loop operator, $W(C) = \prod_{\ell^* \in C} \sigma_{\ell^*}$~\cite{Kogut79}. The correctable regime corresponds to the deconfined (Higgs) phase, where the expectation value of the Wilson loop follows a ``perimeter law,'' $[\langle W(C) \rangle] \sim \exp(-P)$. 
This suppresses the large magnetic flux tubes (thermal fluctuations) that would cause logical errors. In contrast, the uncorrectable regime is the confinement phase, which exhibits an ``area law,'' $[\langle W(C) \rangle] \sim \exp(-A)$.

\begin{figure}[t!]
	\centering
	\includegraphics[width=0.7\columnwidth]{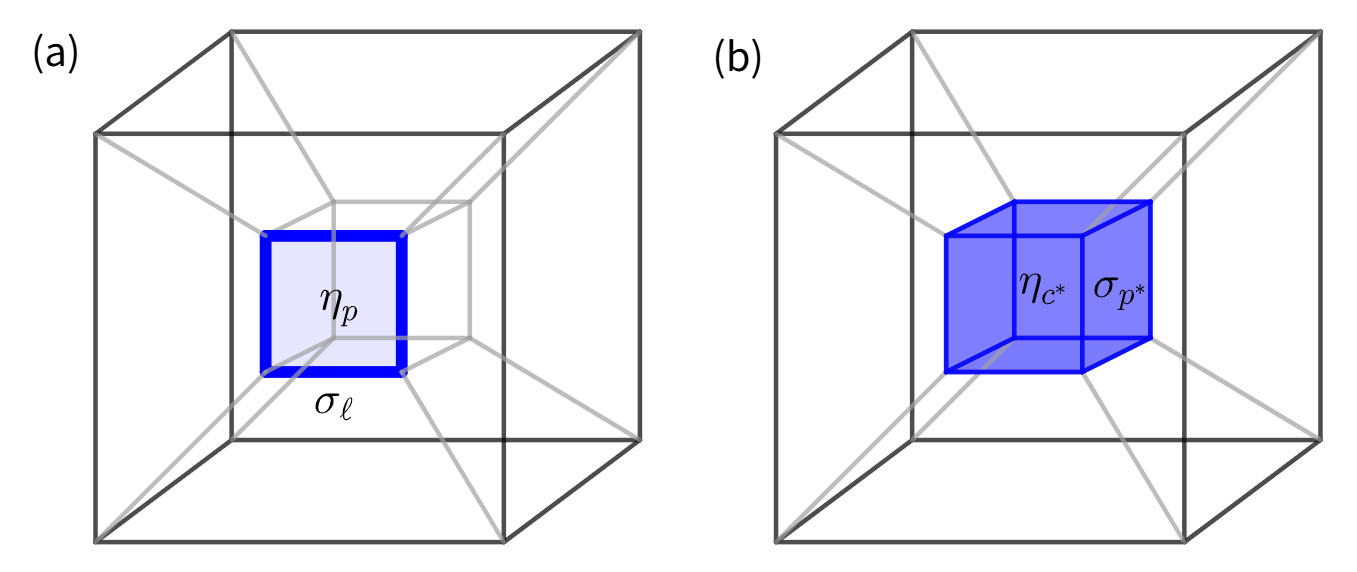}
	\caption{\textbf{Illustration of 4D $\mathbb{Z}_2$ lattice  gauge models on a tesseract.} (a) Random-Plaquette Gauge Model (RPGM): This model is defined by $\mathbb{Z}_2$ spin variables $\sigma_\ell =\pm 1$ residing on the links $\ell$ of the hypercubic lattice. The Hamiltonian involves a four-body interaction on plaquettes $p$, with a random coupling sign $\eta_p =\pm 1$. (b) 
    Random-Cube Gauge Model (RCGM): This model is defined on the dual lattice. The spin variables $\sigma_{p^*}=\pm 1$ reside on each plaquette $p^*$, and a six-body interaction with random coupling sign $\eta_{c^*}=\pm 1$ is located on each 3D cubes $c^*$.}
	\label{lattice_4D}
\end{figure}

\textbf{Imperfect Measurements (4D Models).}
When measurements are faulty, the analysis is elevated to a 4D spacetime lattice. The fundamental unit of this lattice is a tesseract (Fig.~\ref{lattice_4D}), and the connectivity and dual relationships are modified accordingly.

For bit-flip ($X$) errors with imperfect measurements, the error chains correspond to 2-chains in a 4D lattice. The cycle constraint maps the problem to the 4D random-plaquette $\mathbb{Z}_2$ gauge model (RPGM), with the Hamiltonian
\begin{equation}
	H^{X,M} = -J \sum_{p} \eta_{p} \prod_{\ell \in \partial p} \sigma_{\ell},
\end{equation}
where spins $\{\sigma_\ell\}$ reside on links and random couplings $\{\eta_p\}$ are on plaquettes. As in the 3D case, the correctable regime corresponds to the deconfined phase, characterized by a perimeter law for the Wilson loop order parameter.
This can be viewed as a straightforward generalization of the Ising gauge theory in high dimensions.

Nevertheless, a crucial distinction is found in the phase-flip error sector with imperfect measurements. 
The $Z$-errors form 1-chains in the 4D lattice, while the homological constraint leads to a 4D 2-form random-cube $\mathbb{Z}_2$ gauge model (RCGM). 
This model is an example of a higher-form, or generalized, lattice gauge theory~\cite{Gaiotto15,Wen19,McGreevy23}, whose properties differ significantly from ordinary gauge theories, as summarized in Table~\ref{Table_higher_form}.

\begin{table}[b]
	\centering
	\caption{\textbf{Comparison between the ordinary and 2-form $\mathbb{Z}_2$ gauge theories.}
    The table summarizes the type of gauge field, symmetry degree, transformation rule, and the nature of their charged objects and Wilson operators.}
	\renewcommand{\arraystretch}{1.3}
	\begin{tabular}{@{}lll@{}}
    \toprule
    \textbf{Aspect} & \textbf{$\mathbb{Z}_2$ gauge theory (ordinary)} & \textbf{$\mathbb{Z}_2^{(1)}$ Gauge Theory (2-form)} \\ \midrule
    Gauge field type & 1-form gauge field $A$ & 2-form gauge field $B$ \\
    Gauge symmetry degree & 0-form $\lambda$ (acts on points) & 1-form $\Lambda$ (acts on lines) \\
    Gauge transformation & $A \to A + \delta\lambda$ & $B \to B + \delta\Lambda$ \\
    Charged objects & Particles (0-branes) & Strings (1-branes) \\
    Wilson operator &
    \makecell[l]{Loop operator \\ $W(C)=\exp\!\left(i\pi\!\int_C A\right)$}
    &
    \makecell[l]{Surface operator  \\ $W(S)=\exp\!\left(i\pi\!\int_S B\right)$} \\
    \bottomrule
\end{tabular}
	\label{Table_higher_form}
\end{table}

The RCGM is fundamentally different from the standard RPGM. While ordinary gauge theories feature a 1-form gauge field (like $A_\mu$) whose sources are point-like particles, a 2-form gauge theory describes a 2-form field whose sources are extended, string-like objects. As detailed in Table~\ref{Table_higher_form}, the gauge symmetry itself is a 1-form, acting on lines rather than points. This elevates the familiar Wilson loop, which measures the flux through a 1D path, to a Wilson surface, which measures the flux through a 2D surface.
Such a membrane operator serves as the order parameter for the phase transition. 

The Hamiltonian for the RCGM is
\begin{equation}
	H^{Z,M} = -J \sum_{c^*} \eta_{c^*} \prod_{p^* \in \partial c^*} \sigma_{p^*}.
\end{equation}
In this model, the spin variables $\{\sigma_{p^*}\}$ reside on plaquettes, while the random couplings $\{\eta_{c^*}\}$ are located on the 3D cubes, defining a six-body interaction, as illustrated in Fig.~\ref{lattice_4D}.  
The disorder-free limit, $\eta_{c^*} \equiv 1$, reduces to the 2-form gauge theory proposed from a pure symmetry construction~\cite{Wegner71,Johnston14}, but here it naturally emerges from the 3D toric code.
The corresponding phase transition is characterized by the Wilson surface operator, $W(S) = \prod_{p^* \in S} \sigma_{p^*}$, where $S$ is a closed 2D membrane. The asymptotic behavior of its average value distinguishes the two phases~\cite{Johnston14}:
\begin{equation}
	[\langle W(S) \rangle] \sim \begin{cases} \exp(-V) & \text{(Deconfined} \leftrightarrow \text{Ordered)} \\ \exp(-A) & \text{(Confined}  \leftrightarrow \text{Disordered)} \end{cases},
\end{equation}
where $V$ is the minimal enclosed volume and $A$ is the area of the surface $S$. The correctable regime for the toric code corresponds to the deconfinement phase, where the ``volume law'' cost suppresses the large-scale thermal fluctuations that cause logical errors.

\begin{figure}[t]
	\centering
	\includegraphics[width=0.7\columnwidth]{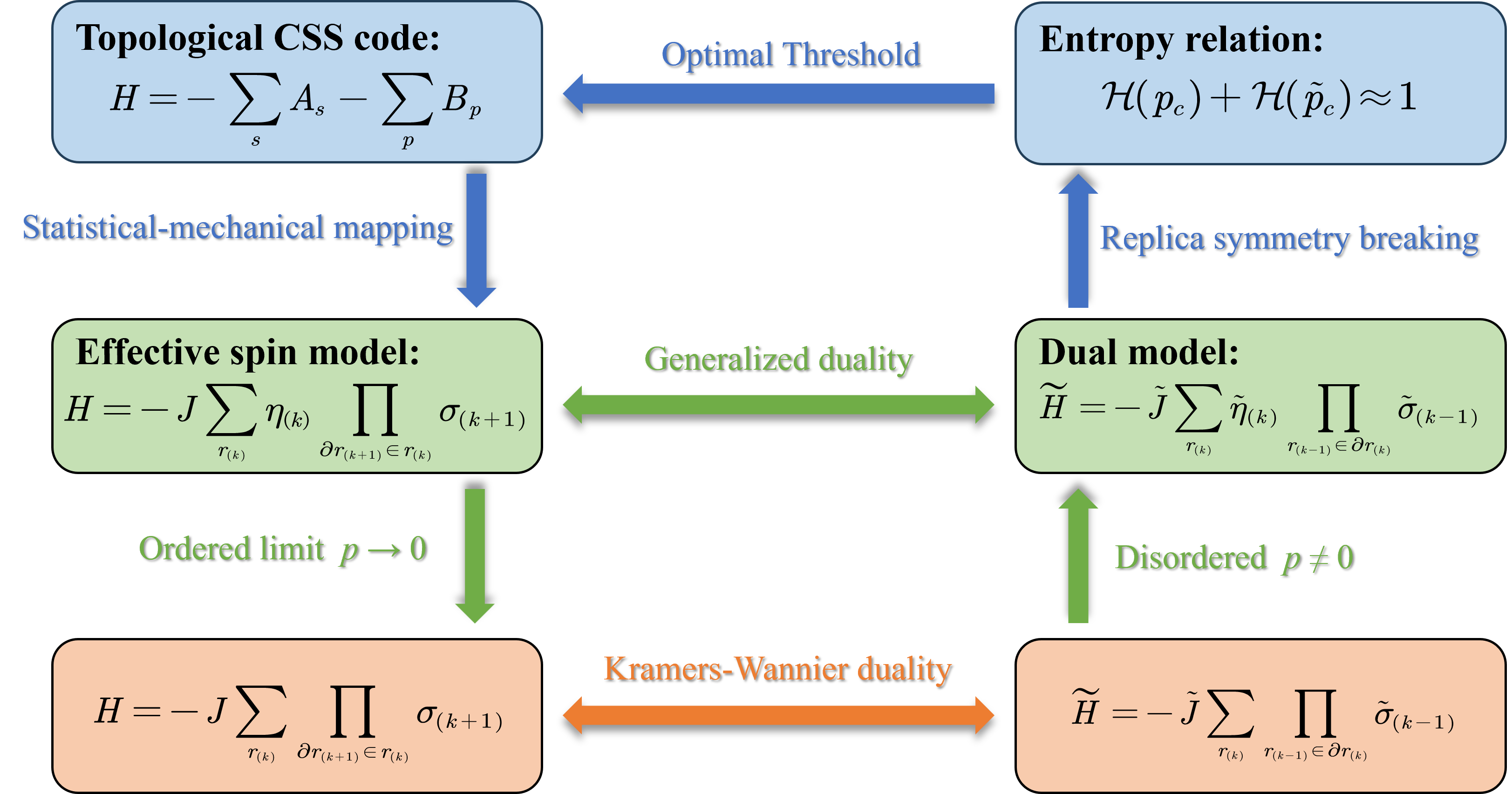}
	\caption{\textbf{Connections between topological codes, statistical mechanics, and thresholds.} This schematic outlines the statistical-mechanical mapping of a topological CSS code to its effective and dual spin models, linked by generalized and Kramers-Wannier dualities, and ultimately to optimal thresholds through an entropy relation.}
	\label{duality_threshold}
\end{figure}

\begin{table}[b]
    \renewcommand{\arraystretch}{1.5}
    \centering
    \caption{\textbf{Effective spin models and optimal thresholds of the 3D Toric Code.}
    The critical points, $p_{c}$ and $\tilde{p}_{c}$, for each model and its dual model satisfy a generalized Shannon entropy relation $\mathcal{H}(p_{c}) + \mathcal{H}(\tilde{p}_{c}) \approx 1$. The optimal thresholds are given by $p_{th} = p_{c}$. References refer to the known numerical estimates of the thresholds. Results with measurement errors are in bold face.}
    \begin{tabular}{|c|c|c|c|}
        \hline
        \hline
        \textbf{Error} & \textbf{Measurement} & \makecell{\textbf{Statistical Mechanical Model} \\ (Dual Model)} & \makecell{$p_{c}$ \\ ($\tilde{p}_{c}$)} \\
        \hline
        $X$ & Perfect & \makecell{3D random-bond Ising model \\ (3D random-plaquette $\mathbb{Z}_2$ gauge model)} & \makecell{$23.3\%$~\cite{Ozeki98} \\ ($3.3\%$~\cite{Ohno2004})} \\
        \hline
        $Z$ & Perfect & \makecell{3D random-plaquette $\mathbb{Z}_2$ gauge model \\ (3D random-bond Ising model)} & \makecell{$3.3\%$~\cite{Ohno2004} \\ ($23.3\%$~\cite{Ozeki98})} \\
        \hline
        $X$ & Faulty & \makecell{{\bf 4D random-plaquette $\mathbb{Z}_2$ gauge model} \\ (Self-dual)} & \makecell{$\mathbf{11\%}$~\cite{Arakawa05} \\ ($11\%$)} \\
        \hline
        $Z$ & Faulty & \makecell{{\bf 4D 2-form random-cube $\mathbb{Z}_2$ gauge model} \\ (4D random-bond Ising model)} & \makecell{ $\mathbf{2\%}$ (This work) \\ ($28\%$~\cite{Hartmann01})} \\
        \hline
        \hline
    \end{tabular}
    \label{tab_threshold}
\end{table}

\section{Optimal thresholds}

In quantum error correction, the error threshold ($p_{th}$) is a critical value of the physical error rate $p$. 
When $p < p_{th}$, the probability of a logical error can be suppressed exponentially by increasing the code size, allowing one to reach an arbitrarily small logical error rate.
For $p > p_{th}$, error correction fails, and encoded information is inevitably lost. The threshold's value is therefore of immense physical significance, as it quantifies the minimum hardware quality required for a given code architecture to be viable. 
Our analysis determines the optimal thresholds, which are the theoretical upper bounds for all decoders under the same noise model.

The procedure to determine the optimal thresholds is summarized in Fig.~\ref{duality_threshold}.
In particular, we leverage a generalized entropy duality, Eq.~\eqref{eq:duality} that extends the standard Kramers-Wannier duality to incorporate quenched disorder.
This generalized duality method relates the critical points of two mutually dual spin models along the Nishimori line, allowing the thresholds to be inferred.

When measurements are perfect, there is no extra time direction, hence the random spin models $H^X$ and $H^Z$ live on the code lattice and its dual lattice, respectively.
The two models associate with the excitation maps~\cite{Haah13} of the quantum code and are mutually dual, namely, $H^Z \equiv \widetilde{H}^X$.
Hence, one can directly plug their thresholds $p^X_{th}$ and $p^Z_{th}$, known as code capacity, into the entropy relation Eq.~\eqref{eq:duality}.

When measurement errors are considered, syndrome propagations in the $X$ sector and $Z$ sector are subject to independent dynamics.
Consequently, the corresponding spin models $H^{X,M}$ and $H^{Z,M}$ are no longer dual to each other.
Nevertheless, we can separately derive a dual model for each of them, as indicated in Table~\ref{tab_threshold}.
The entropy relation Eq.~\eqref{eq:duality} holds for $H^{X,M}$ and its corresponding dual $\widetilde{H}^{X,M}$; similar for $H^{Z,M}$ and $\widetilde{H}^{Z,M}$.
Therefore, one can always analyze the phase transition of the simpler model within a dual pair $H$ and $\widetilde{H}$, and subsequently infer the transition point of the more complex model.

The thresholds are summarized in Table~\ref{tab_threshold}.
In the case of perfect measurements, the threshold for bit-flip ($X$) errors is $p_{th}^X \approx 23.3\%$, while for phase-flip ($Z$) errors, it is $p_{th}^Z \approx 3.3\%$.
These values are known and determined by high-precision simulations of the 3D random-bond Ising model (RBIM) and its dual, the 3D random-plaquette $\mathbb{Z}_2$ gauge model (RPGM)~\cite{Wang03}.

For bit-flip errors with faulty measurements, the corresponding spin model is the 4D random-plaquette gauge model, which is self-dual.
In this case, the duality relation provides direct predictive power.
The self-duality forces the phase transition to occur at the Shannon entropy $\mathcal{H} \approx 1/2$, by which one immediately has $p_{th}^{X,M} \approx 11\%$~\cite{Nishimori07}.

The effective spin model corresponding to phase-flip errors with faulty measurements is the 4D 2-form random-cube gauge model.
There are no existing numerics for this model, as simulating complex gauge theories is generally difficult and lacks efficient algorithms.
Nonetheless, according to the generalized duality, this 2-form random gauge theory is dual to a 4D random-bond Ising model.
The phase transition of the 4D random Ising model is known from Monte Carlo simulations, which exhibits a critical point at $\tilde{p}_c \approx 28\%$~\cite{Hartmann01}.
Therefore, without resorting to expensive simulations of the more exotic gauge theory, leveraging the duality relation Eq.~\eqref{eq:duality}, we can infer that the phase-flip threshold under faulty measurement is $p^{Z,M}_{th} \approx 2\%$.

The overall phenomenological threshold is given by the lower values of the two, hence  $p^{Z,M}_{th} \approx 2\%$.
This threshold is considerably high considering the complex stabilizer structures of the 3D toric code.

The significant asymmetry between $p^{Z,M}_{th}$ and $p^{X,M}_{th}$ reflects the distinction of their logical operators.
As shown in Fig.~\ref{logical_operators}, the logical $X$ operator is a 2D membrane, whose weight is quadratic to the linear lattice size.
Thus, the code can tolerate more $X$-errors.
From a phase transition perspective, it indicates that the deconfined phase of the standard Ising gauge theory is more robust than the $2$-form one.
We conjecture such kind of large asymmetry in thresholds to be a ubiquitous feature for 3D codes, with exception of self-dual fracton codes~\cite{Vijay16,Canossa24}.
This also implies rooms to improve the resilience of 3D codes by engineering biased $X$ and $Z$ error channels~\cite{Pablo21,Huang23}.

\section{Conclusion and Discussion}

In this work, we studied the fault-tolerance threshold of the 3D toric code, a paradigm for 3D topological codes, under imperfect syndrome measurements.
Accounting for measurement error is essential, as it is an unavoidable source of error in realistic quantum devices and fundamentally alters a code's performance.
We derived the effective 4D random gauge models that describe the code's correctability and computed their thresholds analytically using duality techniques.
Our calculations reveal that the 3D toric code has phenomenological thresholds $p^Z_{th} \approx 0.02$ for phase-flip errors and $p^X_{th} \approx 0.11$, modestly reduced from the values in the absence of measurement errors.  
This finding underscores the inherent robustness of the 3D toric code and constitutes a substantial advance compared to previous studies of 3D topological codes.
Furthermore, the thresholds we determined are optimal, hence can also be used to benchmark the development of new decoders.

Beyond its immediate relevance to quantum error correction, our work introduces a novel 4D random 2-form $Z_2$ gauge theory with interdisciplinary interest.
Higher-form gauge theories have recently become a research frontier in both high-energy and condensed matter physics, as they provide an essential language for describing extended excitations and classifying exotic topological states of matter~\cite{Gaiotto15,Wen19,McGreevy23}. 
While the disorder-free limit of this random 2-form $Z_2$ gauge theory has been constructed elsewhere for high-energy motivation~\cite{Johnston14}, the natural emergence of its disorder-full version from a QEC code is a notable finding. 
This shows that complex quantum codes serve as a fertile ground for discovering new gauge theories, illustrating the cross-disciplinary nature of research into fault-tolerant quantum computing.

The statistical-mechanical mapping approach and duality techniques have a plausible history in analyzing the performance of topological codes and continue to garner significant attention.
It is instructive to compare the methodology used in this work with other recently developed methods, such as the information-diagnosis approach~\cite{Fan24,Li21}. 
This latter approach shares the core principle of relating a code's correctability to a phase transition in a statistical-mechanical model. 
However, rather than working directly with the density matrix $\rho$ and von Neumann entropy, it relates the $n$-th moment of $\rho$  to the $n$-th R\'{e}nyi entropy $S^{(n)}=\frac{1}{1-n} \log \left(\operatorname{tr} \rho^n\right)$.
This procedure maps the problem onto a $(n-1)$-flavor spin model with $n\geq 2$.
The thresholds obtained in this manner establish an upper bound on the optimal threshold, whereas our approach yields the exact value. 
Nonetheless, the information-diagnosis method provides intuitive correspondence between physical quantities on both sides of the mapping. 
For instance, the quantum relative entropy on the code side can be related to correlation functions in the effective spin model, while mutual information can be interpreted as the free energy penalty of domain walls along non-trivial loops~\cite{Lyons24,Su24}.

Our methodology is also applicable to other higher-dimensional topological codes~\cite{Bombin07,Haah11,Vijay16} and topological qLDPC codes~\cite{Chen25,Liang25,Roeck25}; most of their optimal thresholds in the presence of imperfect measurements remain unknown. As rapid hardware advancements bring the physical realization of these complex codes within reach, understanding their fault-tolerance thresholds becomes increasingly critical. The richer structures of these codes will inevitably lead to more complicated effective statistical models and gauge theories. While such theories are interesting in their own right and may reveal new many-body physics and phases of matter, determining their phase transitions directly is expected to be difficult. Nevertheless, the generalized duality relation can shift this problem to a more tractable dual description. For instance, the phenomenological error models of 3D color codes and fracton codes are expected to give rise to novel higher-form gauge theories in four dimensions. Their duals may be 4D local spin models or subsystem-symmetric models that are practically more accessible. Furthermore, this framework has the potential to be extended to circuit-level error models, thereby incorporating correlated noise and error propagation. Such an extension would account for the full complexity of realistic noise processes and enable a more direct assessment of circuit-level implementations of quantum codes.

\ack{This work is supported by the National Natural Science Foundation of China (Grants No.~12522502, No.~12474145, and No.~12447101), the Strategic Priority Research Program of Chinese Academy of Sciences (Grant No.~XDB1680000), 
Shanghai Municipal Science and Technology Major Project (Grant No.~2019SHZDZX01), Anhui Initiative in Quantum Information Technologies,
the Fundamental Research Funds for the Central Universities
(Grant No.~lzujbky-2024-jdzx06), the Natural Science Foundation of Gansu Province (Grants No.~22JR5RA389 and No.~25JRRA799), and the “111 Center” under Grant No.~B20063.
MAMD acknowledges support from Spanish MICIN grant PID2021-122547NB-I00 and the ``MADQuantumCM'' project funded by Comunidad de Madrid (Programa de acciones complementarias) and by the Ministry for Digital Transformation and of Civil Service of the Spanish Government through the QUANTUM ENIA project call –Quantum Spain project, and by the European Union through the Recovery, Transformation and Resilience Plan Next Generation EU within the framework of the Digital Spain 2026 Agenda, the CAM Programa TEC-2024/COM-84 QUITEMAD-CM.}




\suppdata{Supplementary Material contains details of the homological formalism, the statistical-mechanical mapping, and the generalized duality method.}

\bibliographystyle{iopart-num}
\bibliography{3D_toric.bib}

\newpage

\end{comment}

\begin{center}
	\textbf{\large --- Supplementary Material ---\\[0.5em]Phenomenological Noise Models and Optimal Thresholds of the 3D Toric Code}\\[1em]
\end{center}

\setcounter{equation}{0}
\setcounter{figure}{0}
\setcounter{table}{0}
\setcounter{page}{1}
\setcounter{section}{0}
\renewcommand{\theequation}{S\arabic{equation}}
\renewcommand{\thefigure}{S\arabic{figure}}
\renewcommand{\thetable}{S\arabic{table}}
\renewcommand{\thesection}{S.\Roman{section}}

\section{Homological formalism for error correction}

\subsection{Error channels and correction}

We first consider noise errors and the process of recovering information from them under perfect measurement. In this discussion, we focus on two types of Pauli error channels affecting the physical qubits in the code: phase-flip ($Z$) errors and bit-flip ($X$) errors, each occurring independently on every qubit with probability $p$. As for bit-phase-flip ($Y$) errors, they can be decomposed into $Y=\mathrm{i}XZ$,  which represents the simultaneous occurrence of both $X$ and $Z$ errors on the same qubit.

It is useful to define the total error channel $E_Z$ (or $E _X$), which is the product of all $Z$ (or $X$) error channels:
\begin{equation}
	E _Z=\prod _{\ell\in\mathcal{E} _Z}Z_{\ell},\qquad E _X=\prod _{\ell\in\mathcal{E} _X}X_{\ell}, 
\end{equation}
where $\mathcal{E} _Z$ ($\mathcal{E} _X$) is the set of all links where the qubits experience $Z$ ($X$) errors. The destruction of the encoded state is described by
\begin{equation}
	\begin{aligned}
		\rho &\to\rho '_Z=E _Z\rho E _Z^\dagger,  \\
		\rho &\to\rho '_X=E _X\rho E _X^\dagger.
	\end{aligned}
\end{equation}
To restore $\rho '$ to $\rho $ after the error, we need to determine the recovery channel $R\in\bar{E}=\{g\}\cdot E$ such that $R\rho 'R^\dagger=\rho $, where we omit the subscript $Z$ or $X$ in the general discussion. It can be seen that the recovery channel $R$ is not unique for a fixed error channel $E$, and we say that these $R$ and $E$ belong to an equivalence class $\bar{E}$ \cite{Bombin12}.

However, this result holds only for a specific error channel. In practice, the only information about errors that we can extract from measurements is the locations of the check operators with eigenvalues $-1$, known as the measurement syndrome $S$, which can be caused by different error channels, and therefore we need to derive an equivalence class from $S$ by the decoding process. We assume the most likely equivalence class derived from the measurement syndrome is $\bar{E}$, which is corrected by the recovery channel $R$. Consequently, we can express the effect of the errors on the density matrix $\rho $ after the measurement as
\begin{equation}
	\rho '={\rm pr}(\bar{E}|S)\bar{E}\rho \bar{E}+\sum_{\bar{E}_L}{\rm pr}(\bar{E}_L|S)\bar{E}_L\rho \bar{E}_L, 
\end{equation}
where ${\rm pr}(\bar{E}|S)$ is the conditional probability that the decoding result is $\bar{E}$ when the measurement syndrome is $S$, and $\bar{E}_L$ are the different equivalence classes from $\bar{E}$.

By the definition of equivalence class, we have $\bar{E}_L=\bar{E}\cdot L$, where $L$ is a Pauli operator. The error channels $E\cdot L$ and $E$ share the same measurement syndrome, yet they do not belong to the same equivalence class, which implies that $L$ commutes with all stabilizers and must be a logical operator, i.e., the equivalence classes differ from each other by logical operations. Therefore, we hope that the most probable class $\bar{E}$ does not introduce the logical operation. For this reason, $\bar{E}$ is called the trivial equivalence class.

\begin{figure}[h!]
	\centering
	\includegraphics[width=0.7\columnwidth]{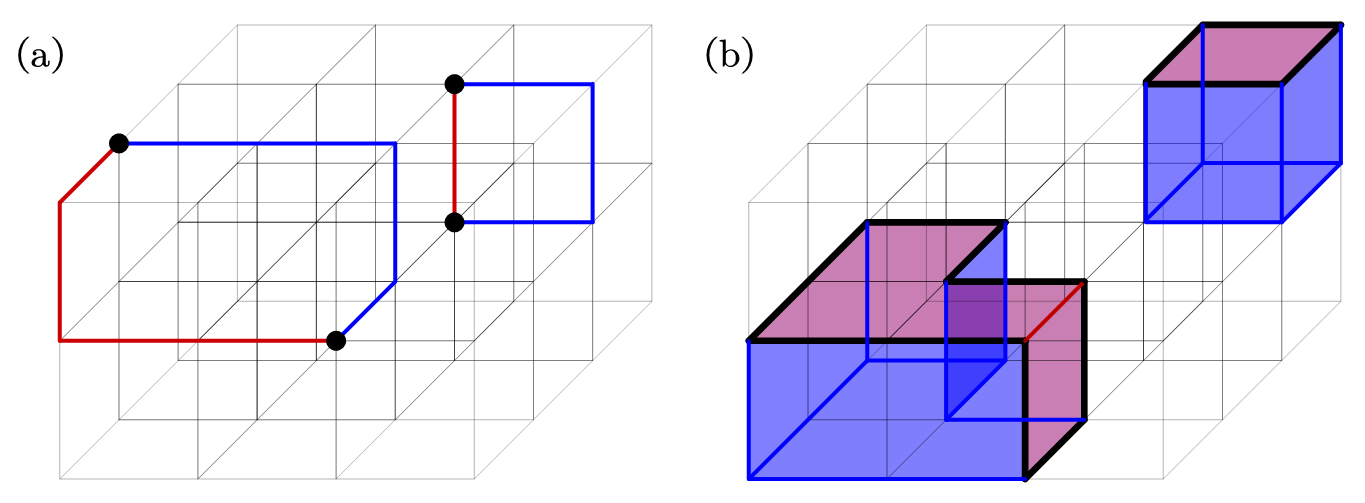}
	\caption{\textbf{Error channel and error correction:} (a) The error channel $E_Z$ (red) is a set of strings in the original lattice, where the syndrome appears at the endpoints of the strings, which is corrected by the non-unique recovery channel $R_Z$ (blue) that makes $R_ZE_Z\in\{g_z\}$ a set of trivial closed strings. (b) The error channel $E_X$ (red) is a set of membranes in the dual lattice, where the syndrome appears at the boundary of the membranes, which is corrected by the non-unique recovery channel $R_X$ (blue) that makes $R_XE_X\in\{g_x\}$ a set of trivial closed membranes.}
	\label{sm_error_channel}
\end{figure}

The error channels and error correction can be intuitively understood from a geometric picture. As shown in Fig.\,\ref{sm_error_channel}(a), for the $Z$ errors, the error channel $E_Z$ consists of a set of strings in the original lattice, with the syndrome appearing at their endpoints. To correct the errors, we need to identify the recovery channel $R_Z$ such that $R_ZE_Z\in\{g_z\}$ is a set of trivial closed strings. For the $X$ errors, as shown in Fig.\,\ref{sm_error_channel}(b), the error channel $E_X$ consists of a set of membranes in the dual lattice, with the syndrome appearing as closed strings at the boundaries of the membranes. Then $X$ errors are corrected by $R_X$ and $R_XE_X\in\{g_x\}$ is a set of trivial closed membranes.

\subsection{Chain complex and homology}

A mapping that assigns an element of $\mathbb{Z} _2=\{0,1\}$ to each object of the same dimension in the lattice is referred to as a $k$-chain $\mathcal{C}^{(k)}$, where $k$ is the dimension of the objects. Without causing confusion, we will use $k$-chains to refer specifically to the set of all objects assigned the value 1 by such a mapping. For instance, when the objects are sites or links, the corresponding maps are the 0-chain and 1-chain, respectively, and similarly for higher-order chains. There are two operations we need to use: addition ``+'' and the differential ``$\partial$''. The $k$-chains with the same dimension can be added to each other, where the values on the same object are summed according to the rules of addition in $\mathbb{Z} _2$, forming a module over $\mathbb{Z} _2$; the boundary operator $\partial$ maps a $k$-chain to its boundary, a $(k-1)$-chain, and satisfies the property $\partial\circ\partial=0$. A chain with a trivial boundary is called a cycle, and there are two distinct types of cycle: a cycle with a trivial topology is said to be homologically trivial, meaning it can be written as the boundary of a higher-order chain; otherwise, it is homologically nontrivial.

In this language, Pauli operators are mapped to the $k$-chains, and the product of operators corresponds to the sum of chains. We use the calligraphic symbols to represent the chains to distinguish them from operators. Thus, in the original lattice, the stabilizer chains $\mathcal{G} _z,~\mathcal{G} _x$ are 2-chains and 0-chains, respectively, with 1 assigned to the locations where check operators exist, while the error chains $\mathcal{E} $ are 1-chains. This allows us to define the chain complex for $Z$ errors
\begin{equation}
	\mathcal{G} _z\xrightarrow{\partial}\mathcal{E} \xrightarrow{\partial}\mathcal{G} _x.
\end{equation}
Here $\mathcal{G} $ and $\mathcal{E} $ in the chain complex represent the sets of all chains, while the specific chains are referred to when they appear alone. In the dual lattice, $\mathcal{G} _z^*,~\mathcal{G} _x^*,~\mathcal{E}^*$ are 1-chains, 3-chains, and 2-chains, respectively, and we have the dual chain complex for $X$ errors
\begin{equation}
	\mathcal{G} _z^*\xleftarrow{\partial^*}\mathcal{E}^* \xleftarrow{\partial^*}\mathcal{G} _x^*.
\end{equation}

The condition $\partial\circ\partial=0$ implies that $\partial\mathcal{G} _z$ ($\partial\mathcal{G} _x^*$) are homologically trivial 1-cycles (2-cycles) in the original (dual) lattice, which agrees with the previous conclusion. Furthermore, the logical operator chains $\bar{\mathcal{Z} }$ ($\bar{\mathcal{X} }$) are homologically nontrivial 1-cycles (2-cycles) in the original (dual) lattice. The equivalence class $\bar{\mathcal{E} }=\mathcal{E} +\{\mathcal{C} \}$ means the chains in it all have the same boundary and the sum of any two chains is a homologically trivial cycle, which is called a homology class.

\subsection{Measurement errors}

In fact, in addition to the physical qubit errors, errors also occur in the syndrome measurement, and the locations of these measurement errors are referred to as ``ghost charges''. If ghost charges are mistaken for genuine charges caused by qubit errors, the error correction procedure described above may introduce additional errors. Fortunately, the effects of measurement errors can be eliminated through repeated measurements. In the absence of measurement errors, error correction can be regarded as finding a recovery operator $R$ that combines with the error channel $E$ to form a trivial loop. This conclusion remains valid even in the presence of measurement errors; merely the description is adjusted. In the following, we will use the framework of chain complexes and homology to analyze them.

First, let us consider the correction of given error chains. For the $Z$ errors, repeated $A_s$ measurements produce a measurement syndrome, represented as a sequence of eigenvalues on the sites of the original lattice along the timeline. To visualize this, we extend the 3D spatial lattice to the 4D spacetime lattice and assign the measurement syndrome to the timelike links, forming a 1-chain $\mathcal{S} _A$, where the links with eigenvalue $-1$ of the check operators are assigned 1. Similarly, the 1-chain $\mathcal{E} _A$ corresponding to the measurement errors is defined, where the timelike links with errors are assigned 1. 

\begin{figure}[t]
	\centering
	\includegraphics[width=0.5\columnwidth]{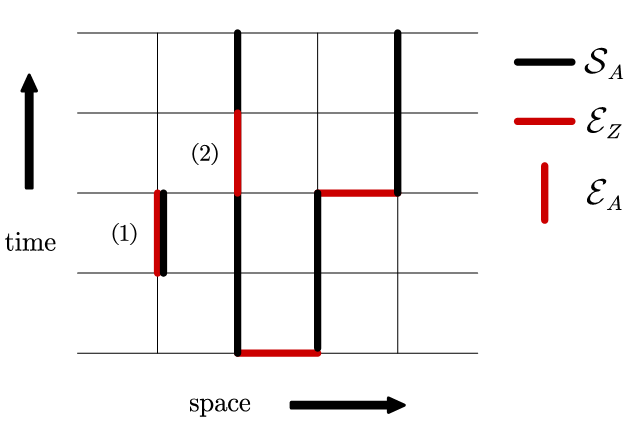}
	\caption{\textbf{World line of the charges:} The red line is the qubit error chain $\mathcal{E} _Z$ and the measurement error chain $\mathcal{E} _A$, and the black line is measurement syndrome chain $\mathcal{S}_A $, where the horizontal chain is the spatial part of the charge's world line and the vertical chain is the temporal part. (1) The measurement error chain coinciding with the syndrome chain counteracts the world line of the ghost charges, and (2) the rest of the measurement error chain integrates the world line of the genuine charges, so that $\mathcal{E}_Z +\mathcal{E} _A+\mathcal{S} _{A}$ is the world line of the genuine charges.}
	\label{sm_world_line}
\end{figure}

In the four-dimensional space-time lattice, the propagation path of charges is their world line, with the sum of the qubit error chain and the syndrome chain $\mathcal{E} _Z+\mathcal{S} _A$ representing the world line of both genuine charges and ghost charges. Nevertheless, the world line of ghost charges is given by the error chain $\mathcal{E} _{A}$, so $\mathcal{E} _Z+\mathcal{E} _{A}+\mathcal{S} _A$ counteracts the ghost charges and turns into the world line of the genuine charges (see Fig.\,\ref{sm_world_line}). The errors are then corrected by the recovery chain $\mathcal{R} $ so that the final propagation path $\mathcal{C} =\mathcal{E} _Z+\mathcal{E} _{A} +\mathcal{S}_A+\mathcal{R}$ forms a homologically trivial 1-cycle in the original lattice, which means the genuine charges vanish. Analogously, we consider the $X$ errors and $B_{\ell^*}$ measurement errors in the four-dimensional dual lattice. In this case, the measurement syndrome chain $\mathcal{S} _B$ and the measurement error chain $\mathcal{E} _B$ are 2-chains. The combination $\mathcal{E} _X+\mathcal{E} _B+\mathcal{S} _B$ counteracts the ghost loops and represents the world sheet swept by the genuine charges. Consequently, the errors are corrected by $\mathcal{R}$ such that $\mathcal{C} =\mathcal{E} _X+\mathcal{E} _{B}+\mathcal{S}_B+\mathcal{R}$ is a homologically trivial 2-cycle in the dual lattice.

Through the above discussion, we have derived a unified correction condition for both types of errors:
\begin{equation}
	\mathcal{E} +\mathcal{S} +\mathcal{R} =\mathcal{C}, \label{correction_condition}
\end{equation}
where $\mathcal{E} $ is the total error chain, including both qubit and measurement errors, and $\mathcal{C} $ is an arbitrary homologically trivial cycle. The main distinction lies in the fact that the relevant chains are 1-chains in the original lattice for $Z$ errors and 2-chains in the dual lattice for $X$ errors. Furthermore, this condition extends and applies to both perfect and imperfect measurements.

In practice, the decoder's goal is to derive an equivalence class $\bar{\mathcal{E} }=\mathcal{S} +\mathcal{R} +\{\mathcal{C} \}$ from the measurement syndrome $\mathcal{S}$, so that the errors are corrected by the recovery chain $\mathcal{R}\in\bar{\mathcal{E} } + \mathcal{S}$. We expect there exists a threshold $p_{th}$ for the error probability of a physical qubit, such that for $p<p_{th}$ and in the limit of $L\to \infty$, the errors can always be successfully corrected, which requires
\begin{equation}
	\lim_{L\to \infty}\frac{{\rm pr}(\bar{\mathcal{E} }_\mathcal{L} |\mathcal{S} )}{{\rm pr}(\bar{\mathcal{E} }|\mathcal{S} )}\to 0, \label{sm_th_condition}
\end{equation}
where $\bar{\mathcal{E} }_\mathcal{L}=\bar{\mathcal{E} } + \mathcal{L}$ is another equivalence class with the logical operation $L$, and $\mathcal{L}$ is a homologically nontrivial cycle carrying a logical operation. 

\section{Derivation of statistical-mechanical mapping}

In the previous section, we derived the error correction condition (\ref{correction_condition}) and the threshold condition (\ref{sm_th_condition}). In the following, we will introduce the general statistical-mechanical mapping for topological codes and demonstrate that the threshold $p_{th}$ corresponds to the critical phase transition point along the Nishimori line in the statistical-mechanical (SM) model.

First, we take $\mathcal{E} $ to be the most probable error chain within the homology class $\bar{\mathcal{E} }$. Accordingly, $\mathcal{E}$ is uniquely determined by the syndrome $\mathcal{S}$ through decoding, so that $\mathcal{S}$ can be represented by $\mathcal{E}$, and we obtain
\begin{equation}
	{\rm pr}(\bar{\mathcal{E} }|\mathcal{S} )=\sum_{\mathcal{E} '\in\bar{\mathcal{E}}}{\rm pr}(\mathcal{E} '|\mathcal{S} )=\sum_{\mathcal{C} }{\rm pr}(\mathcal{E} +\mathcal{C} |\mathcal{E}  ).
\end{equation}
Without loss of generality, we assume physical qubits are put on the links of the lattice with a probability $p$ of experiencing an error, so the probability of $\mathcal{E} $ occurring is
\begin{equation}
	{\rm pr}(\mathcal{E})=\prod_\ell(1-p)^{1-n_\mathcal{E}(\ell)}p^{n_\mathcal{E}(\ell)}=\left[\prod_\ell(1-p)\right]\cdot\prod_\ell\left(\frac{p}{1-p}\right)^{n_\mathcal{E}(\ell)}, 
\end{equation}
where 
\begin{equation}
	n_\mathcal{E}(\ell)=\begin{cases}~1,&\mathrm{if}~\ell\in\mathcal{E} \\~0,&\mathrm{if}~\ell\notin\mathcal{E} \end{cases}.
\end{equation}
When measurement syndrome $\mathcal{S} $ is observed, the conditional probability of error chain $\mathcal{E} '=\mathcal{E} +\mathcal{C} \in\bar{\mathcal{E} }$ occurring is expressed as
\begin{equation}
	{\rm pr}(\mathcal{E} +\mathcal{C}|\mathcal{E} )=\left[\prod_\ell(1-p)\right]\cdot\prod_\ell\left(\frac{p}{1-p}\right)^{n_{\mathcal{E} +\mathcal{C}}(\ell)}=\left[\prod_{\ell}\sqrt{p(1-p)}\right]\cdot\prod_{\ell}\exp\big(\beta J\eta_{\ell}u_{\ell}\big),
\end{equation}
where $n_{\mathcal{E} +\mathcal{C}}(\ell)=n_{\mathcal{E}}(\ell)+n_{\mathcal{C}}(\ell)-2n_{\mathcal{E}}(\ell)n_{\mathcal{C}}(\ell),~\eta_{\ell}=1-2n_\mathcal{E}(\ell),~u_{\ell}=1-2n_\mathcal{C}(\ell)$, and
\begin{equation}
	e^{-2\beta J}=\frac{p}{1-p}
\end{equation}
is the Nishimori line.

Now we can denote an SM model with the formal Hamiltonian
\begin{equation}
	H=-J\sum_{\ell}\eta_{\ell}u_{\ell}, \label{formal_Hamiltonian}
\end{equation}
with the partition function $Z[J,\eta]=\sum_{u_{\ell}}\exp(-\beta H)$. Here the parameters $\{\eta _\ell\}$ are determined by a given error chain $\mathcal{E} $, while the configurations of statistical-mechanical variables $\{u_\ell\}$ represent the cycle $\mathcal{C} =\mathcal{E} +\mathcal{E} '$. Thus, we have
\begin{equation}
	{\rm pr}(\bar{\mathcal{E} }|\mathcal{S} )=\sum_{\mathcal{C} }{\rm pr}(\mathcal{E} +\mathcal{C} |\mathcal{E}  )\propto Z[J,\eta]
\end{equation}
and ${\rm pr}(\bar{\mathcal{E} }_\mathcal{L}|\mathcal{E} )\propto Z_L[J,\eta]$ for the nontrivial homology class $\bar{\mathcal{E} }_\mathcal{L}$. Therefore, the relative probabilities of homology classes can be expressed as
\begin{equation}
	\frac{{\rm pr}(\bar{\mathcal{E} }_\mathcal{L} |\mathcal{S} )}{{\rm pr}(\bar{\mathcal{E} }|\mathcal{S} )}=\frac{Z_L[J,\eta]}{Z[J,\eta]}=\exp[-\beta \delta F(J,\eta)], \label{energy_difference}
\end{equation}
where $\delta F(J,\eta)=F_L(J,\eta)-F(J,\eta)$ is the free energy difference between the homology classes.

In the SM model, the cycles $\mathcal{C} $ and $\mathcal{L} $ correspond to thermal fluctuations in the form of closed loops with $u_\ell=-1$. These thermal fluctuations arise from the cost of free energy, which is proportional to the length of the closed loops. In the limit of $L\to \infty$, the length of the nontrivial cycle $\mathcal{L} $ tends to infinity, resulting in a divergent free energy cost and thus $ F_L(J,\eta)\to\infty$. When $p<p_{th}$, Eqs.\,(\ref{sm_th_condition}) and (\ref{energy_difference}) imply $ \delta F(J,\eta)\to \infty$, causing $F(J,\eta)$ to converge to a constant; conversely, $ \delta F(J,\eta) $ converges while $F(J,\eta)$ diverges for $p>p_{th}$. This indicates that the success or failure of error correction corresponds to two distinct phases of the SM model, with the threshold $p_{th}$ corresponding to the phase transition critical point along the Nishimori line. When the errors are correctable, $F(J,\eta)$ converges, infinite thermal fluctuations are absent, and this is referred to as the ordered phase; otherwise, $F(J,\eta)$ diverges, infinite thermal fluctuations may arise, which is called the disordered phase, where the errors are uncorrectable.

The above conclusion is valid for a specific error chain $\mathcal{E}$, corresponding to a particular error correction process. To obtain a threshold applicable to all possible error correction events, we treat $\eta_{\ell}$ as a random parameter with the probability distribution
\begin{equation}
	P(\eta_{\ell})=p\delta (\eta_{\ell}+1)+(1-p)\delta (\eta_{\ell}-1), 
\end{equation}
which makes the SM model a random model. Then we perform a configuration average to calculate the quenched free energy
\begin{equation}
	\big[F(J,\eta)\big] =\int\prod_{\ell}d\eta_{\ell}P(\eta_{\ell})F(J,\eta), \label{quenched_energy}
\end{equation}
and use it instead of $F(J,\eta)$ \cite{BookNishimori}. Significantly, the random model may include a third phase, known as the spin-glass phase. However, since the phase transition under consideration occurs along the Nishimori line, which has been shown not to pass through the spin-glass phase \cite{Song22}, the latter is excluded from our discussion.

Finally, for the formal Hamiltonian (\ref{formal_Hamiltonian}), the difference in the expression of $u$ comes from its cycle conditions. For example, for the 3D toric code, $u$ satisfies
\begin{equation}
	\prod _{\partial \ell\in s}u_{\ell}=1, \quad \text{or} \quad \prod _{\partial p^*\in \ell^*}u_{p^*}=1,
\end{equation}
such that $\mathcal{C} $ is a 1-cycle in the original lattice or a 2-cycle in the dual lattice.

\section{Derivation of duality method}

\subsection{Exact Kramers-Wannier duality}

To illustrate the generalized duality method, we first present the exact Kramers-Wannier duality. In the ordered limit $p=0$, the Hamiltonian (\ref{formal_Hamiltonian}) is expressed as
\begin{equation}
	H=-J\sum_{\ell}u_{\ell}, 
\end{equation}
and we consider it satisfying $\prod _{\partial \ell\in s}u_{\ell}=1$ to be the 3D plaquette $\mathbb{Z} _2$ gauge model. Following Wegner's construction technique for the duality \cite{Wegner73}, we write the partition function as
\begin{equation}
	Z=N_g\sum_{\{u_{\ell}\}} \prod_s \delta\left(\prod_{\ell \in \partial s} u_\ell, 1\right) \prod_\ell \omega_\ell(u_\ell), 
\end{equation}
where $\omega_\ell(u_\ell)=e^{\beta Ju_\ell}$ is the Boltzmann factor on a link $\ell$, $N_g=2^{N-1}$ is ground-state degeneracy. Then we expand the delta function constraint using a set of dual spin variables $\{\tilde{\sigma}_s\}$ located on the sites of the dual lattice, and obtain
\begin{equation}
	\begin{aligned}
		Z&=N_g\sum_{\{u_{\ell}=\pm1\}}\prod _{s}\left(\frac12\sum_{\tilde{\sigma }_i=\pm1}\prod _{\partial \ell\in s}\tilde{\sigma }_i^{(1-u_\ell)/2}\right)\prod _{\ell}\omega_\ell(u_\ell) \\
		&=\frac12\sum_{\{\tilde{\sigma }_i=\pm1\}}\prod _{\ell}\left(\sum_{u_{\ell}=\pm1}\omega_\ell(u_\ell)\prod_{s\in\partial \ell}\tilde{\sigma }_i^{(1-u_\ell)/2}\right) \\
		&=\frac12\sum_{\{\tilde{\sigma }_i=\pm1\}}\prod _{\ell}\left(e^{\beta J}+e^{-\beta J}\tilde{\sigma }_i\tilde{\sigma }_j\right)
	\end{aligned}, \label{partition_plaquette}
\end{equation}
where the product subscript $\partial \ell\in s$ is for $\ell$, while $ s\in\partial \ell$ is for $s$.

Since the form of the dual spin interaction in Eq.\,(\ref{partition_plaquette}) is $\tilde{\sigma }_i\tilde{\sigma }_j$, we define the 3D Ising model as the dual model
\begin{equation}
	\widetilde{Z}=\sum_{\{\tilde{\sigma }_i=\pm1\}}\prod _{\langle ij\rangle }\exp\big(\tilde{\beta }\tilde{J}\tilde{\sigma }_i\tilde{\sigma }_j\big), 
\end{equation}
where $\langle ij\rangle$ represents the link $\ell$. Comparing two equations, we obtain the duality relation
\begin{equation}
	\sinh(2\beta J)\sinh(2\tilde{\beta }\tilde{J})=1, \label{duality_relation}
\end{equation}
and
\begin{equation}
	Z=2^{N/2-1}(\sinh2\tilde{\beta }\tilde{J})^{-N/2}\widetilde{Z}.
\end{equation}
Consequently, we arrive at the result of the exact Kramers-Wannier duality: the 3D plaquette $\mathbb{Z} _2$ gauge model is dual to the 3D Ising model, and they are precisely the SM models for the $Z$ errors and $X$ errors in the ordered limit $p=0$.

In general, we can write $u_\ell$ as $u_{(k)}$, which is defined on $r_{(k)}$ and satisfies the restricted condition $\prod _{\partial r_{(k)}\in r_{(k-1)}}u_{(k)}=1$, where $r_{(k)}$ represents site $s$, link $\ell$, plaquette $p$, or cube $c$. Thus we obtain
\begin{equation}
	H=-J\sum_{r_{(k)}}\prod_{\partial r_{(k+1)}\in r_{(k)}}\sigma_{(k+1)}, 
\end{equation}
and
\begin{equation}
	\begin{aligned}
		Z&=2^{-N_{(k-1)}}N_g\sum_{\{\tilde{\sigma }_{(k-1)}\}}\prod _{r_{(k)}}\bigg(e^{\beta J}+e^{-\beta J}\prod _{r_{(k-1)}\in\partial r_{(k)}}\tilde{\sigma }_{(k-1)}\bigg) \\[5pt]
		&=2^{N_{(k)}/2-N_{(k-1)}}N_g(\sinh2\tilde{\beta }\tilde{J})^{-N_{(k)}/2}\widetilde{Z}
	\end{aligned}, \label{partition_general}
\end{equation}
where $N_{(k)}$ is the number of $r_{(k)}$ in the code, $\tilde{\sigma }_{(k-1)}$ is the dual spin on $r_{(k-1)}$. Therefore, the Hamiltonian of the dual model is
\begin{equation}
	\widetilde{H}=-\tilde{J}\sum_{r_{(k)}}\prod _{r_{(k-1)}\in\partial r_{(k)}}\tilde{\sigma }_{(k-1)}.\label{duality_Hamiltonian}
\end{equation}
In this case, the duality relation (\ref{duality_relation}) remains valid.

\subsection{Generalized duality}

Next we consider the generalized duality relation for $p\neq 0$. In order to calculate the quenched free energy (\ref{quenched_energy}), we use the relation
\begin{equation}
	F=-\frac{1}{\beta }\ln Z=-\frac{1}{\beta }\lim_{n\to0}\frac{Z^n-1}n, 
\end{equation}
which reduces the problem to calculating $[Z^n]$ and taking the limit $n\to 0$. This approach, known as the $n$-replica method, involves preparing $n$ replicas of the original system and evaluating the configuration average of the product of the partition functions.

According to Eq.\,(\ref{partition_plaquette}) and the first equation in (\ref{partition_general}), the $n$-replica partition function can be expressed as
\begin{equation}
	\begin{aligned}
		\big[Z^n\big]&=\Bigg[\frac{N_g^n}{2^{nN_{(k-1)}}}\sum_{\{\tilde{\sigma }_{(k-1)}^i\}}\prod _{r_{(k)}}\prod _{i=1}^n\bigg(e^{\beta J\eta_{(k)}}+e^{-\beta J\eta_{(k)}}\prod _{r_{(k-1)}\in\partial r_{(k)}}\tilde{\sigma }_{(k-1)}^i\bigg)\Bigg] \\
		&=\frac{N_g^n}{2^{nN_{(k-1)}}}\sum_{\{\tilde{\sigma }_{(k-1)}^i\}}\prod _{r_{(k)}}\left(\sum_{\mathbf{u}_{(k)}}\omega _{(k)}(\mathbf{u}_{(k)})\prod _{i=1}^n\prod _{r_{(k-1)}\in\partial r_{(k)}}\tilde{\sigma }_{(k-1)}^{i~(1-u_{(k)})/2}\right)
	\end{aligned}, \label{replica_partition}
\end{equation}
where $[\cdots] = \int\prod_{(k)}d\eta_{(k)}P(\eta_{(k)})\cdots$ represents the configuration average, $\omega _{(k)}(\mathbf{u}_{(k)})=pe^{-(n-2|\mathbf{u}_{(k)}|)\beta J}+(1-p)e^{(n-2|\mathbf{u}_{(k)}|)\beta J}$ is the disorder-averaged Boltzmann factor, $\mathbf{u}_{(k)}=(u_{{(k)}}^1,u_{{(k)}}^2,\cdots,u_{{(k)}}^n)$ is an $n$-dimensional vector with components $u_{{(k)}}^i = \pm1$, and $|\mathbf{u}_{(k)}|$ is the number of $u_{{(k)}}^i=-1$ components in $\mathbf{u}_{(k)}$. As shown in Ref.\,\cite{Jean-Marie}, Eq.\,(\ref{replica_partition}) can be simplified as
\begin{equation}
	\left[Z^n\right]=2^{nN_{(k)}-nN_{(k-1)}}N_g^n(\cosh \beta J)^{nN_{(k)}}\sum_{\{\tilde{\sigma }_{(k-1)}^i\}}\prod _{r_{(k)}}w_{(k)}(\tilde{\mathbf{u}}_{(k)}), 
\end{equation}
where
\begin{equation}
	w_{(k)}(\tilde{\mathbf{u}}_{(k)})=\begin{cases}~(\tanh \beta J)^{|\tilde{\mathbf{u}}_{(k)}|},&|\tilde{\mathbf{u}}_{(k)}|~\text{is even}\\~(\tanh \beta J)^{|\tilde{\mathbf{u}}_{(k)}|+1},&|\tilde{\mathbf{u}}_{(k)}|~\text{is odd},  \end{cases}
\end{equation}
and $\tilde{u}_{{(k)}}^i=\prod _{r_{(k-1)}\in\partial r_{(k)}}\tilde{\sigma }_{(k-1)}^i,~\tilde{\mathbf{u}}_{(k)}=(\tilde{u}_{{(k)}}^1,\tilde{u}_{{(k)}}^2,\cdots,\tilde{u}_{{(k)}}^n)$ is the dual variable of $\mathbf{u}_{(k)}$.

On the other hand, we can denote Hamiltonian (\ref{duality_Hamiltonian}) with $\tilde{p}\neq 0$ ($\tilde{p}$ is a new parameter independent of $p$) as the dual model and obtain
\begin{equation}
	\begin{aligned}
		\left[\widetilde{Z} ^n\right]&=\left[ \sum_{\{\tilde{\sigma }_{(k-1)}^i\}}\prod _{i=1}^n\exp\bigg(\tilde{\beta }\tilde{J}\sum_{r_{(k)}}\tilde{\eta }_{(k)}\prod _{r_{(k-1)}\in\partial r_{(k)}}\tilde{\sigma }_{(k-1)}^i\bigg)\right] \\
		&=\sum_{\{\tilde{\sigma }_{(k-1)}^i\}}\prod _{r_{(k)}}\tilde{\omega} _{(k)}(\tilde{\mathbf{u}}_{(k)})
	\end{aligned}, 
\end{equation}
where $\tilde{\omega} _{(k)}(\tilde{\mathbf{u}}_{(k)})=\tilde{p}e^{-(n-2|\tilde{\mathbf{u}}_{(k)}|)\tilde{\beta }\tilde{J}}+(1-\tilde{p})e^{(n-2|\tilde{\mathbf{u}}_{(k)}|)\tilde{\beta }\tilde{J}}$, and $\tilde{\beta }\tilde{J}$ is given by the Nishimori line with $\tilde{p}$.

It is observed that the $n$-replica partition functions $[Z^n]$ and $[\widetilde{Z}^n]$ no longer follow the exact Kramers-Wannier duality relation. Nevertheless, a generalized duality relation can be established at the critical point along the Nishimori line \cite{Nishimori07}. We rewrite $[\widetilde{Z}^n]$ as
\begin{equation}
	\left[\widetilde{Z}^n\right]=\big(\tilde{\omega}(0)\big)^{N_{(k)}}\sum_{\{\tilde{\sigma }_{(k-1)}^i\}}\prod _{r_{(k)}}\tilde{w} _{(k)}(\tilde{\mathbf{u}}_{(k)}), 
\end{equation}
where $\tilde{\omega}(0)=\tilde{p}e^{-n\tilde{\beta }\tilde{J}}+(1-\tilde{p})e^{n\tilde{\beta }\tilde{J}}$, and
\begin{equation}
	\tilde{w} _{(k)}(\tilde{\mathbf{u}}_{(k)})=\frac{\tilde{\omega} _{(k)}(\tilde{\mathbf{u}}_{(k)})}{\tilde{\omega}(0)}=\frac{e^{(n+1-2|\tilde{\mathbf{u}}_{(k)}|)\tilde{\beta }\tilde{J}}+e^{-(n+1-2|\tilde{\mathbf{u}}_{(k)}|)\tilde{\beta }\tilde{J}}}{e^{(n+1)\tilde{\beta }\tilde{J}}+e^{-(n+1)\tilde{\beta }\tilde{J}}}.
\end{equation}
By definition, $\tilde{w} _{(k)}(\tilde{\mathbf{u}}_{(k)},\tilde{\beta }\tilde{J})$ as a function of the variable $\tilde{\beta }\tilde{J}$ decreases monotonically from 1 to 0, while $w _{(k)}(\tilde{\mathbf{u}}_{(k)},\beta J)$ for $\beta J$ increases monotonically from 0 to 1. Since the phase transition is unique, we expect
\begin{equation}
	w _{(k)}(\tilde{\mathbf{u}}_{(k)},\beta J)\approx \tilde{w} _{(k)}(\tilde{\mathbf{u}}_{(k)},\tilde{\beta }\tilde{J})
\end{equation}
near the critical point. Accordingly, we obtain the approximate equation
\begin{equation}
	\left[Z^n\right]\approx 2^{nN_{(k)}-nN_{(k-1)}}N_g^n(\cosh \beta J)^{nN_{(k)}}\big(\tilde{\omega}(0)\big)^{-N_{(k)}}\left[\widetilde{Z} ^n\right].
\end{equation}

The above derivation goes from $[Z^n]$ to $[\widetilde{Z}^n]$; analogously, we can also perform the same procedure from $[\widetilde{Z}^n]$ to $[Z^n]$ to get $ \tilde{w} _{(k)}(\mathbf{u}_{(k)},\tilde{\beta }\tilde{J})\approx w_{(k)}(\mathbf{u}_{(k)},\beta J)$ and
\begin{equation}
	\left[\widetilde{Z} ^n\right]\approx 2^{nN_{(k)}-nN_{(k+1)}}\tilde{N}_g^n(\cosh \tilde{\beta }\tilde{J})^{nN_{(k)}}\big(\omega(0)\big)^{-N_{(k)}}\left[Z^n\right].
\end{equation}
Combining two equations and using the identity $N_g\tilde{N}_g=2^{N_{(k-1)}+N_{(k+1)}-N_{(k)}}$ \cite{Wegner71}, we have
\begin{equation}
	\frac{\omega(0)\tilde{\omega}(0)}{(\cosh \beta J)^n(\cosh \tilde{\beta }\tilde{J})^n}\approx 2^n.
\end{equation}
In the limit of $n\to0$, taking the logarithm of the above equation will give the approximate entropy relation
\begin{equation}
	\mathcal{H}(p_c)+\mathcal{H}(\tilde{p}_c)\approx 1,
\end{equation}
where $\mathcal{H}(p)=-p\log_{2}(p)-(1-p)\log_{2}(1-p)$ is the Shannon entropy.


\end{document}